\documentclass[11pt, oneside]{article}
\usepackage[T1]{fontenc}
\usepackage[english]{babel}

\usepackage[a4paper, margin=1in]{geometry}
\usepackage{setspace}
\usepackage{parskip}

\usepackage{fancyhdr}

\usepackage{enumitem}
\usepackage{array}
\usepackage{booktabs}
\usepackage{caption}
\usepackage{longtable}
\usepackage{tabularx}
\usepackage{multirow}
\usepackage{tikz}
\usetikzlibrary{arrows.meta,decorations.pathmorphing}

\usepackage{microtype} 

\usepackage{graphicx}
\usepackage{float}
\usepackage{pgf-pie}

\usepackage{ragged2e}
\usepackage{sectsty}
\usepackage[autostyle]{csquotes} 
\renewcommand{\mktextdel}[1]{[\textellipsis]} 
\SetBlockThreshold{0} 
\usepackage{hyperref}

\usepackage[style=apa, backend=biber]{biblatex} 
\title{The Future of Scholarly Blogs: Scholarly Bloggers’ Perspectives on Long-Term Preservation}
\date{}

\begin{document}

\author{
Catharina Ochsner\thanks{Berlin School of Library and Information Science, Humboldt-Universität zu Berlin, Unter den Linden 6, 10099 Berlin, Germany.
Email: \href{mailto:catharina.ochsner@hu-berlin.de}{catharina.ochsner@hu-berlin.de}.\textbf{ORCID}: \href{https://orcid.org/0009-0005-3885-3951}{0009-0005-3885-3951}}\\
{\small IBI, HU Berlin}
 \and 
Heinz Pampel\thanks{Berlin School of Library and Information Science, Humboldt-Universität zu Berlin, Unter den Linden 6, 10099 Berlin, Germany and Helmholtz Association, Helmholtz Open Science Office, Telegrafenberg, 14473 Potsdam, Germany.
Email: \href{mailto:heinz.pampel@hu.berlin.de}{heinz.pampel@hu-berlin.de}.\textbf{ORCID}: \href{https://orcid.org/0000-0003-3334-2771}{0000-0003-3334-2771}}\\
{\small IBI, HU Berlin \& Helmholtz}
}

\maketitle

\pagestyle{fancy}
\fancyhf{}
\fancyhead[R]{\thepage}
\section*{Abstract}
Scholarly blogs are an important venue for scholarly communication, however they are not yet integrated into the preservation workflows of digital research and information infrastructures, which places their long-term access at risk. By using Star and Ruhleder’s (\citeyear{star1996}) dimensions of information infrastructure as a theoretical framework, in this study we investigate social and organizational factors that impact blog preservation and derive implications for the implementation of blogs into an information infrastructure. We conducted and qualitatively analyzed 13 semi-structured interviews with scholarly bloggers to identify bloggers' perceived challenges concerning blog preservation and their requirements an information infrastructure tasked at preserving scholarly blogs. We found that participants named or indirectly described existing infrastructure institutions and their services that are already working towards preserving the scholarly record. The findings suggest extending the scope of existing infrastructure institutions to include scholarly blog preservation.
\newpage

\pagenumbering{arabic}

\section{Introduction}
Within the broad landscape of scientific information dissemination, scholarly blogs are a distinct type of scholarly publication, that facilitate the communication of research both within scientific communities and to wider audiences \autocite{luzon2013}. Compared to more formal types of publications, such as journal articles, conference proceedings, or monographs, blogs are accessible and low-cost for both scholars and readers and enable fast, informal, and open communication \autocite{burton2015, fenner2022}. As part of the scholarly record, more formal types of scholarly publications are routinely integrated into digital research and information infrastructures \autocite{borgman2007}, meaning the shared distribution of social, organizational, and technical systems and activities that enable the support of research practices \autocite{bowker2010} that include preservation activities \autocite{borgman2007}. As a more informal and fleeting publication type, scholarly blogs are not comprehensively integrated into information infrastructures and therefore not preserved \autocite{ochsner2025bfp}. As a consequence, scholarly blogs are at risk of being lost and leaving a gap in the scholarly record \autocite{ochsner2025jdoc}. However, as scholarly objects, scholarly blogs need to be identifiable by reference systems and be accessible in the long-term to give readers the opportunity to trace back and verify sources \autocite{eve2024}.

While the existing but limited research on the topic of blog preservation has found that individual bloggers, libraries, and platform providers have made efforts to make scholarly blogs more accessible and citable as scholarly output (e.g., through their inclusion in library catalogs, the assignment of identifiers and licenses), these efforts remain scarce and do not ensure the long-term preservation of blogs \autocite{ochsner2025jdoc}. Additionally, prior research has discerned that while bloggers are generally interested in the preservation of their blogs, they rely on personal backups instead of infrastructure institutions \autocite{arangodocio2011, ochsner2026iconf}. Proposed solutions for blog preservation have been of technical nature on the integration of blogs into digital libraries \autocite{limani2017}, or web archival strategies for blogs which included a software for the aggregation of blogs \autocite{kalb2013}. However, none of these technical recommendations have been widely implemented by infrastructure institutions.

This article enhances the existing technical contributions, by focusing on social and organizational factors of blog preservation. To do this, we conducted 13 semi-structured interviews with German scholarly bloggers to identify bloggers' challenges and requirements concerning blog preservation to derive implications for their implementation into an information infrastructure. This study is guided by the following research questions (RQs):

\begin{table}[H]   \captionsetup{justification=centering, format=plain}
    \centering
    \begin{tabular}{p{1.0\linewidth}}
    \textbf{RQ 1:} What challenges do scholarly bloggers encounter in engaging with blog preservation?\\
    \textbf{RQ 2:} Considering these challenges, according to scholarly bloggers...\\
    \hspace{2em}\textbf{RQ 2.1.} ... what blog content should be preserved by an information infrastructure?\\
    \hspace{2em}\textbf{RQ 2.2.} ... what services do scholarly bloggers’ require from an information infrastructure?\\
    \hspace{2em}\textbf{RQ 2.3.} ... how could an information infrastructure for scholarly blogs be organized and governed?\\
    \end{tabular}
\end{table}

This study proceeds with a review of the relevant literature and a description of our method. We then present our findings, discuss their broader conceptual implications, and derive suggestions for relevant stakeholders that support blog preservation. While we present timely empirical data on bloggers' challenges and requirements for an information infrastructure tasked with blog preservation, this article also makes a conceptual contribution to the existing research on infrastructure studies, scholarly communication and digital preservation by conceptualizing the results along Star and Ruhleder’s (\citeyear{star1996}) dimensions of information infrastructure.
\section{Literature Review}
\subsection{Blogs - Definition and Background}
Blogs are commonly described as publicly accessible webpages, where bloggers publish diary-style content in the form of blog posts that are listed in reverse chronological order, with the newest post appearing first \autocite{nentwich2012}. Blogs first emerged in the late 1990's \autocite{rettberg2008} and gained more prominence with the emergence of the Web 2.0 in the 2000's \autocite{fenner2008} in which participatory digital environments were increasingly characterized by user-generated content \autocite{nentwich2012}. With the popularization of the Web 2.0, blogs on the subject of science and scholarship emerged \autocite{fenner2008} and increasingly integrated into the scholarly practice \autocite{ferguson2010}. In the following we will call these blogs, scholarly blogs and define them as blogs written by scholars or people with a profession close to research or a background in academia (e.g., science journalists, practitioners, university students, or teachers), about research or research-related topics \autocite{littek2012}. Within the scholarly literature on the topic a variety of other terms (e.g., academic blogs, research blogs, science blogs, scientific blogs) have been used seemingly interchangeably. While science blogs have sometimes been defined to be blogs on the topic of natural sciences \autocite{puschmann2014, blanchard2013, kouper2010}, the literature does not address differences between other terms. Scholars merely distinguish between blogs with a specific purpose or content \autocite{walker2006}, like research program blogs, event blogs, laboratory blogs, thesis blogs, seminar blogs \autocite{koehler2008} and research group blogs \autocite{luzon2006}. Blogs are now used for the internal scientific communication between scholars and the external scholarly communication that also involves stakeholders such as publishers, libraries, and the public \autocite{burton2015, shema2014}. While more formal types of scientific communication are embedded in established preservation systems \autocite{burton2015, fenner2022}, newer digital formats such as blogs operate within infrastructures that are far more fluid and weakly governed. Scholarly blogs lack mechanisms for their registration and preservation, which results in the loss of blog content and consequently gaps the scholarly record \autocite{ochsner2025jdoc}.

\subsection{Information Infrastructures}
Research is increasingly shaped by digital practices and infrastructures that organize and support scholarly communication \autocite{borgman2007}. The term infrastructure has been assigned different meanings by scholars and practitioners \autocite{edwards2024}. Instead of defining technical characteristics that make up an infrastructure, scholars from the science and technology studies (STS) describe infrastructure as a relational concept \autocite{bowker2010, star1996}. Accordingly, infrastructures are not built but grown organically within an already existing environment \autocite{edwards2007, edwards2009}, emerging out of the information practices and needs of a community that are deeply interconnected with their activities and organizational structures \autocite{star1996}. Star and Ruhleder (\citeyear{star1996}) have defined dimensions of infrastructures that are summarized in Table \ref{table:1}.

\begin{table}[H]
    \captionsetup{justification=centering, format=plain}
    \caption{Dimensions of Infrastructure according to Star and Ruhleder (\citeyear{star1996})}
    \label{table:1}
    \centering
    \renewcommand{\arraystretch}{1.1}
    \begin{tabular}{p{0.17\linewidth} p{0.76\linewidth}}
        \toprule
        \textbf{Dimension} & \textbf{Definition} \\
        \midrule
         Embeddedness & Infrastructure is embedded into social, organizational and, technical factors.\vspace{1pt}\\
         Transparency & Infrastructure is transparent to use, does not have to be reinvented for each task and invisibly supports those tasks.\vspace{1pt}\\
         Reach or scope & Infrastructure extends across multiple events and practices.\vspace{1pt}\\
         Learned as part \par of membership & Infrastructure becomes invisible and taken for granted by members of a community of practice. Outsiders see it as something they must learn.\vspace{1pt}\\
         Links with \par conventions of \par practice & Infrastructure shapes and is shaped by the shared conventions and practices of its community.\vspace{1pt}\\
         Embodiment \par of standards & Infrastructure incorporates standards that enable a transparent interoperability with other tools and systems, even amid conflicts.\vspace{1pt}\\
         Built on an \par installed base & Infrastructure evolves by layering new elements onto old ones, creating a dependence on the installed base which both enables progress and constrains innovation.\vspace{1pt}\\
         Becomes visible \par upon breakdown & Normally invisible and working infrastructure becomes visible when it breaks while back-ups highlight now-visible infrastructure.\vspace{1pt}\\ 
        \bottomrule
    \end{tabular}
\end{table}

Taken together, the configuration of the dimensions explained in Table \ref{table:1} form an infrastructure \autocite{star1996}. We explain how information infrastructures can support digital curation and preservation below.

\subsection{Digital Curation and Preservation}
The continuously increasing amount of digital data require efforts to curate, manage and preserve relevant objects, which is enacted through digital curation \autocite{abbott2009, lee2007, poole2016, yakel2007}. While the access to digital resources are primarily ensured by archives and libraries \autocite{borgman2007, nentwich2003}, storing digital content does not guarantee long-term accessibility. As a means to ensure that the scholarly record is preserved indefinitely \autocite{eve2024}, infrastructure institutions practice digital preservation, or the \enquote{[...] series of managed activities necessary to ensure continued access to digital materials for as long as necessary} \autocite{dpc}. In the context of long-term preservation, accessibility means the ability to continuously use an original digital resource with all it's qualities and functionalities intact \autocite{dpc}. Digital preservation can be enacted through web archiving \autocite{lee2002}, which plays a significant role in safeguarding cultural heritage and knowledge. Web archives can either curate objects through broad, non-selective web-crawls or by curating collections built in cooperation with libraries and other memory institutions \autocite{kalb2013}.

\subsubsection{Actors in Blog Preservation}
In previous work we have identified relevant actors in blog preservation. We discuss these actors below.
\paragraph{Libraries} Libraries are responsible for the access, preservation and curation of scholarship and research \autocite{borgman2007}. Especially academic libraries provide access to research outputs for students and researchers of their parent institutions \autocite{corrall2013}. Some national libraries already collect and archive blogs. For example, the German National Library's collection mandate includes online publications \autocite{dnbsammelauftrag}. The library already includes blogs in it's catalog and archival system and assigns International Standard Serial Numbers (ISSNs), an identifier for serial publications, to blogs listed in it's catalog \autocite{dnbissn, ochsner2025jdoc}. Other examples for national libraries that already collect blogs are the \href{https://www.library.gov.au/}{National Library of Australia}, the \href{https://www.bnf.fr/}{Bibliothèque nationale de France}, the \href{https://nsk.hr/en/}{National and University Library in Zagreb}, or the \href{https://bac-lac.on.worldcat.org/discovery}{Bibliothèque et Archives Canada}. Additionally, digital libraries are enhanced information storage and retrieval systems that store data and metadata and enable the creation, search, and use of information \autocite{borgman2007}. This makes digital libraries a combination of research, demonstration, and production systems \autocite{bishop2000}. For example, the Lots Of Copies Keep Stuff Safe (LOCKSS) program is a digital library initiative by the Stanford University Libraries \autocite{lockss_people} that provides open-source technologies and services in the support of digital preservation \autocite{lockss_program}. The LOCKSS concept operates as a foundation for the digital preservation of content in libraries by promoting community-based approaches to digital preservation \autocite{lockss_people, lockss_why}.

\paragraph{Web Archives} The most commonly used web archive is the Internet Archive, a non-profit initiative that aims to create a digital library of web pages. The Internet Archive works with libraries to curate and save relevant web pages and also allows individuals to save web pages \autocite{internetarchive_about, kalb2013}. Another example is the The World Class Digital Preservation Services (CLOCKSS), an international digital archive that preserves diverse online scholarly resources \autocite{clockss_about, clockss_home, clockss_how}. CLOCKSS is a community-led collaboration of academic publishers and research libraries \autocite{clockss_home} and a dark archive for the long-term preservation of digital content. Therefore archived resources are inaccessible to the general public and mostly utilized for restoration or recovery purposes in the event of a disaster, data corruption, or other interruptions \autocite{clockss_about, eve2024}. A concrete effort for the preservation of blogs has been made through Rogue Scholar, a dedicated digital archive for science blogs. The service is based on the open-source repository software InvenioRDM and relies on the structured information supplied via blog feeds to curate and preserve content \autocite{fenner2022, fenner2023}. Furthermore, Rogue Scholar assigns DOIs to blog posts and subscribes to the Internet Archive’s “Archive-It” program, enabling systematic long-term archiving of all participating blogs \autocite{roguescholarwhatisrs}. 

\paragraph{Platform Providers} Bloggers can choose between hosting a blog themselves or the hosting through a platform. Commonly used commercial blogging platforms are \href{https://wordpress.com/de/}{Wordpress.com}, \href{https://substack.com/}{Substack}, \href{https://medium.com/}{Medium}, or \href{https://www.blogger.com/about/?hl=de}{Blogger} A significant share of scholarly blogs are hosted on dedicated scholarly blogging platforms like Hypotheses, SciLogs, and ScienceBlogs \autocite{ochsner2025jdoc}. Hypotheses hosts blogs from the humanities and social sciences and is part of OpenEdition, a digital publishing framework designed to disseminate scholarly content in these fields \autocite{hypotheses} that OpenEdition is affiliated with the \textit{École des Hautes Études en Sciences Sociales (EHESS)} \autocite{EHESS2015}, a French public research institution \autocite{EHESS2015, EHESS2016}. The blogging platform SciLogs is an interdisciplinary platform hosted by the publishing house Spektrum der Wissenschaft (eng: spectrum of science) \citeyear{scilogs2014}), while ScienceBlogs originated as an international platform, with its German branch hosted by the Konradin Media Group \autocite{scienceblogs}. Platform providers can support preservation and citability by providing standardized citation proposals, PID's like DOIs and ISSN, and mechanisms for the incorporation of structured metadata and licensing information \autocite{ochsner2025jdoc}.

\paragraph{Bloggers} In earlier work we found that while most bloggers are interested in the long-term accessibility of blogs, they expressed the desire to opt into preservation and decide what content will be preserved. Bloggers also often rely on personal backups \autocite{ochsner2026iconf}. Bloggers that want to support the long-term access, citability and discoverability of their blogs, can apply open licenses, such as those provided by Creative Commons (CC) \autocite{ochsner2025jdoc}, which specify the terms under which content may be shared or reused \autocite{creativecommons}. The gold standard for open access publications is the CC-BY license \autocite{redhead2012} that enables reuse as long as attribution is given to the author \autocite{creativecommons2019}.
\section{Method}
\paragraph{Semi-structured Interviews} This study aims to identify (1) scholarly bloggers challenges concerning blog preservation and to (2) bloggers' requirements for an information infrastructure for the preservation of scholarly blogs. To answer the research questions, we adopted an exploratory qualitative design using semi-structured interviews that are widely used in qualitative research and particularly suitable for exploring specific phenomena while maintaining flexibility in the interaction with participants \autocite{werner2013}. We developed an interview guide to provide structure, while allowing for adaptation of phrasing and questions, depending on the individual interview context. Each interview began with general questions about the participant’s background, blogging activities, and blogging history, followed by more specific questions related to infrastructure. The interview guide intentionally did not define the terms “information infrastructure” or “preservation”. Instead, participants were encouraged to articulate their own interpretations of these concepts and foreground their perspectives.

\paragraph{Sampling and Participants} We conducted interviews with scholarly bloggers since they are the producers of the content to be preserved and also potential users of a future preservation infrastructure. While other actors, such as libraries, web archives and platform providers play important roles in preservation, this study prioritizes the perspective of those whose practices and needs the infrastructure is intended to support. We followed a sampling strategy aimed at capturing heterogeneity among German scholarly bloggers in order to explore a broad range of experiences until saturation was reached \autocite{werner2013}. We sought diversity in gender identity, disciplinary background, institutional affiliation, and blogging platform. To identify potential participants, we used a dataset of 866 German scholarly blogs \autocite{ochsner2025dataset}. Because the majority of blogs in the dataset (77.48\%) originated from the humanities and social sciences, reflecting broader patterns in scholarly blogging \autocite{voigt2024}, we sought to mirror this disciplinary distribution in our sample. We also included bloggers from different platforms to ensure infrastructural diversity. We aimed to conduct at least twelve interviews, drawing on research suggesting that thematic saturation is often reached around this threshold \autocite{guest2006, guest2020}. Ultimately, we conducted thirteen interviews. To preserve anonymity, each participant was assigned a participant ID (P1–P13). Table \ref{table:2} provides an overview of the participants, the discipline they blogged in, their gender as perceived by the authors, and their career stage according to the European Commissions’ Researchers’ Career Framework that classifies research careers into four stages: First Stage Researchers (R1), Recognized Researcher (R2), Established Researcher (R3), and Leading Researcher (R4) \autocite{europeancommission2011}. The blogging platform of individual participants was omitted to ensure the participants’ anonymity.

\begin{table}[H]
    \captionsetup{justification=centering, format=plain}
    \caption{Participants}
    \label{table:2}
    \centering
    \renewcommand{\arraystretch}{1.1}
    \begin{tabular}{p{0.1\linewidth} p{0.3\linewidth} p{0.1\linewidth} p{0.15\linewidth}}
        \toprule
        \textbf{ID} & \textbf{Discipline} & \textbf{Gender} & \textbf{Career stage}\\
        \midrule
        P1 & Humanities \& Social Sciences & Male & R1\vspace{2pt}\\
        P2 & Humanities \& Social Sciences & Male & R2\vspace{2pt}\\
        P3 & Humanities \& Social Sciences & Male & R4\vspace{2pt}\\
        P4 & Humanities \& Social Sciences & Female & R1\vspace{2pt} \\
        P5 & Natural Sciences & Male & R1\vspace{2pt}\\
        P6 & Life Sciences & Female & R1\vspace{2pt}\\
        P7 & Humanities \& Social Sciences & Female & R1\vspace{2pt}\\
        P8 & Natural Sciences & Male & R2\vspace{2pt}\\
        P9 & Humanities \& Social Sciences & Female & R3\vspace{2pt}\\
        P10 & Humanities \& Social Sciences & Male & R1\vspace{2pt}\\
        P11 & Engineering Sciences & Male & R1\vspace{2pt}\\
        P12 & Humanities \& Social Sciences & Female & R2\vspace{2pt}\\
        P13 & Humanities \& Social Sciences & Male & R4\vspace{2pt}\\
        \bottomrule
    \end{tabular}
\end{table}

\paragraph{Data Collection} To assess the clarity and feasibility of the interview guide, we conducted two pre-tests. The pre-tests lasted 42 and 31 minutes and revealed that participants struggled to describe specific long-term preservation practices when asked overly broad questions. As a result, we refined the guide by incorporating concrete scenarios and examples of preservation practices. Participants were contacted via e-mail and received information about the study’s purpose, procedures, and data protection measures. Before the interview, participants were informed about their rights and provided written consent for participation and for the recording and processing of audio and screen data. Interviews were conducted in German between January 30 and March 26, 2025. Twelve interviews were conducted and recorded via \href{https://zoom.us/de/join}{Zoom} and one via the open-source platform \href{https://bigbluebutton.org/}{BigBlueButton}, according to the participant's preferences. Interviews lasted between 35 and 51 minutes (average: 44 minutes). Audio recordings excluded greetings, introductory study information, explanations of participant rights, and closing remarks, as these sections contained no analytically relevant material. To ensure anonymity, neither the audio recordings nor the transcripts were published.

\paragraph{Data Analysis} All interviews were transcribed using the GDPR-compliant software \href{https://www.amberscript.com/de/}{Amberscript}. Participants were informed in advance about the use of Amberscript and consented accordingly. Each transcript was then reviewed by the authors to ensure accuracy and completeness. Repetitions were omitted only if they did not contribute additional meaning. The transcription was conducted in German and all quotations used in this article were translated into English by the authors. Data analysis followed the qualitative content analysis approach described by Kuckartz and Rädiker (\citeyear{kuckartz2024}). To code the interviews, we used MAXQDA 24, a qualitative research software widely used in Library and Information Science (LIS) research \autocite{marjaei2019}. We did not use any generative Artificial Intelligence (AI) for the data analysis. Instead, the anonymized transcripts were imported into MAXQDA and coded using a structured coding system. Our coding frame consisted of deductive codes derived from the interview guide and inductive codes that emerged directly from the data (e.g., data loss, participants’ prior knowledge, issues of trust). The coding frame was organized hierarchically across multiple levels. All codes were formulated in English to ensure conceptual consistency. All interviews were coded by the first author. The code and reviewed transcripts were reviewed by the second author.
\section{Findings}
\subsection{Challenges in Blog Preservation}
Participants reported a range of personal challenges they perceived have negatively impacted their engagement with blog preservation. Several participants declared that they did not possess knowledge about long-term preservation, preservation practices, and information infrastructures (P6 - P8, P11). Participants often described technical and legal expertise to be a challenge (P2, P4, P8). For example, P4 was concerned about preparing metadata in a manner that would support interoperability and reuse. P2 and P7 reported uncertainty about how data security considerations apply to their blogs, while others highlighted specific issues such as rights management of images or comments (P1, P4, P6). Concerning rights management, participants stated that it would not be feasible to ask every person involved in each blog post to consent to having the content preserved, especially concerning comments. P7 also stated that they worry about not complying with their university’s policies. Participants (P1, P7, P9, P12, P13) reported they time to both maintain their blog and acquire knowledge on preservation. P10 illustrated this challenge by noting: \enquote{Editorial work. We do all of this on the side. And that is a problem for many blogs, that they lack stable funding to actually do editorial work, which is a form of unpaid labor} (P10). P5 and P8 explained that a lack of institutional support also contributes to not having enough time for blogging or blog-related topics. P2 stated that they worry about not having a tenured position and therefore not being able to work on the blog and maintain it in the long term. 

Participants were able to anticipate several possible infrastructural challenges concerning the long-term preservation of scholarly blogs. P1 highlighted the increasing complexity of maintaining interconnected blog environments, noting: 

\blockquote{The more interconnected you want it to be, the more difficult it becomes, because blogs were intended to respond to other blogs. Then you have to consider what kind of data structures we actually store the blogs in, in order for them to be comprehensive and meaningful. (P1)}. 

They further elaborated: \enquote{That means you would actually have to preserve this blog ecosystem in its entirety, and that's pretty difficult} (P1). Another participant emphasized the temporal fragility of digital materials: \enquote{But you also have to think a little bit long-term: what will happen in 50 years? And 50 years is almost an eternity for digitally stored media} (P5). Many participants (P1, P3, P6, P8, P13) additionally acknowledged that not all content can be preserved and that determining which objects are sufficiently relevant constitutes a significant archival challenge. P4 and P8 remarked that interoperability could be a potential challenge and some participants (P3, P4, P6) also identified versioning as a likely challenge and emphasized the need for different versions of a blog post to remain visible within any preservation infrastructure. Related to this, P3 and P6 expressed broader concerns about quality control for blogs and blog posts that might be incorporated into research infrastructures.

\subsection{Requirements for an Information Infrastructure}

\subsubsection{Requirement 1: Content}
Participants use their blogs to share a diverse range of information related to their research and professional interests. For example, participants blogged about a research project (P4), a research object (P7, P8), or their dissertation (P2, P8, P9). However, not all participants use their blogs to disseminate research. Participants who did not use their blogs to publish original research instead used blogs to engage in discourse and evaluation of various research topics (P1, P5, P11), personal scientific interests (P5) science policy (P2, P3), or to publish reviews (P8, P10). Furthermore participants promoted their own research by announcing newly published work (P1, P11, P13). Although participants expressed a general desire for their blogs to remain long-term accessible, they articulated differing views on which types of content should be preserved. Some participants regarded all materials as worthy of preservation, as illustrated by P13: \enquote{As long as there is enough storage space, I would save everything} (P13). At the same time, they acknowledged the limitations of preservation infrastructures, noting: \enquote{That's an archival problem that has to be sorted out} (P13). 

Other participants (P3, P6, P8) similarly emphasized that not all blog posts are sufficiently relevant to warrant preservation. As one participant noted: \enquote{So, first of all, I would like to raise the question of whether it is really necessary to save everything?} (P1). Several participants expressed a desire for bloggers to have agency in determining which materials should be preserved: \enquote{If you could control that individually, that wouldn't be bad. Let the bloggers decide} (P11). When asked what elements of their blog and blog posts should be saved, praticipants named texts (P2, P5, P7, P9 - P13), pictures (P2, P3, P6, P7, P9, P11, P13), comments (P2 - P4, P6, P9 - P13), layout elements (P1, P3, P6, P11), audio files (P7), and references or links (P7, P9 - P11). Conversely, participants identified several elements to be not relevant for preservation, including posts that announce calls for papers, outdated posts, or posts not related to scientific output. P3 explicitly stated that images were not important to them, while P7 claimed that the layout was not significant to them.

Some bloggers also distinguished between preserving blogs for the purpose of safeguarding their content and preserving blogs as potential research objects. According to participants, the latter would necessitate capturing the broader blogosphere at the network and citation level, rather than focusing solely on individual posts. One participant articulated this distinction as follows: 

\blockquote{If it's just a matter of understanding the content, blog posts are sufficient. But if you want to analyze blogs as a media phenomenon or as a socio-technical construct, that's not enough. Then you need to have the context. You actually need the form of presentation, because you have to be able to understand how people received or could have received these things at the time. So it [preservation] really depends very much on what you presuppose as future usage intentions. (P1)}

\subsubsection{Requirement 2: Services}
Most participants did not propose concrete technical and organizational requirements and instead articulated broad ideas about what long-term preservation should entail. One participant emphasized several key expectations: 

\blockquote{It's always been crucial that things are reliably available in the long-term, that URLs are stable and available in the long-term, and that things can be cited, especially when they are cited in print formats. If you cite a link that can no longer be found a year later, the whole thing somehow loses its meaning. This means that there has always been an expectation that an institution would simply maintain stability in this regard. (P2)}

P11 emphasized the importance of the provision of DOIs for blog posts. With respect to functional requirements for a preservation infrastructure, P6 said that they want it to be searchable by date and that the operators of this infrastructure provide editorial services such as updating posts that are not up to date, organizing peer reviews, making sure bloggers are institutionally affiliated and sorting out content that is not relevant. P7 stated it was important to them that content is not changed retrospectively. They also expressed a desire for transparency regarding the conditions under which content would be made accessible in the long term, emphasizing the need to trust that these conditions would be upheld. Furthermore, they wished for guidance on technical knowledge, such as data handling and storage, IT literacy, but also licensing and correctly using images. Another participant emphasized the importance of metadata and metadata literacy. They went on to say: \enquote{It [metadata] affects workflows and means more steps. At the same time scholarly blogs that cannot acquire this knowledge should be provided with the structures and perhaps tools and automation, to make it as easy as possible for them.} (P4). One participant, expressing concerns about the long-term stability of digital content and suggested producing physical copies of blogs as an additional preservation strategy. As they explained: 

\blockquote{We'll have to see how storage media develops over the next few decades. How long it will remain stored, or whether people will actually sit down and print out my blog, for example. Bind the whole thing, put two or three versions of it somewhere, just so that it physically exists. That seems to me to be a rather important point with texts like these, that you have to get away from the digital a little bit. (P5)}

\subsubsection{Requirement 3: Governance}
A prominent concern for several participants was the question of governance. Participants reflected on which types of institutions they considered sufficiently trustworthy to take responsibility for preserving blogs. P13 argued that this responsibility should lie exclusively with government-funded institutions, stating: \enquote{For me things like cultural heritage are not tasks of private institutions} (P13). P13 explained that they did not trust commercial actors such as Google as a company or Google services. P13 also went on to express concerns over cyber attacks on their blog or an infrastructure that preserves blogs. Other participants expressed concerns about public institutions. For example, P7 reported not trusting libraries to ensure long-term preservation, whereas P6 expressed a high level of trust in libraries. Another participant described the precarity of the situation as follows: \

\blockquote{I think we need a mix of different players involved. Because one thing is clear from the current political situation, especially when you look at the US right now, with the Trump administration and Elon Musk. Social media and internet platforms are finding themselves in a strange situation where there are deliberate attempts to censor scientific language. We need to protect ourselves from that and that means it actually has to have a very strong base in the scientific community itself. If I were to apply that to the situation in Germany, I would say that, to a certain extent, it has to be the German Research Foundation. But I'm also seeing in the US how the National Science Foundation (NSF) is being accessed and regulated. [...] I also believe that we should not rely on national solutions in this area, but international ones, so that no single government can encroach on such infrastructures. (P3)}

Several participants supported these concern about broader political developments. P10 noted that the political situation in the United States worried them and that they feared similar dynamics emerging in Germany. P12 further described hesitations about blogging on controversial or politically sensitive topics. At the same time, some participants identified a countervailing potential in scholarly blogs to resist anti-science developments. As one participant articulated: 

\blockquote{We need to become politically resilient with all these digital platforms and blogs. I no longer consider academic freedom to be guaranteed. I think we need to equip ourselves somehow to develop resilience in this area. And I believe that blogs will play a very important role here, blogs that may not even be hosted on your own national platform, but somewhere internationally. (P3)}

Participants that had concrete ideas or wishes concerning governance, discussed the following stakeholders that were previously introduced in section 2, as institutions that could preserve scholarly blogs:

\paragraph{Libraries} Some participants (P5 - P7, P13) identified libraries as key stakeholders responsible for ensuring the long-term preservation of their blogs, mentioning institutions such as the German National Library (P5, P13) or university libraries (P7). P10 and P13 expressed a desire for their blog posts to be automatically captured either by the German National Library as part of a mandatory deposit process, while P9 wanted their university library to include their blog into it's catalogs. \enquote{At the latest when a blog is assigned an ISSN, the library should normally also preserve this publication in the long-term, i.e., the National Library} (P13). The participant further suggested: \enquote{I would think that it's a task for the German National Library that would then accordingly need positions and require a reorganization maybe toward digital publications} (P13). Lastly, participants named several additional actors that are often afiliated with libraries that they considered relevant for securing scholarly blogs, including Fachinformationsdienste (specialist information services, P12), subject-specific repositories (P4), and Europeana or other international or European solutions (P3, P11, P13).

\paragraph{Web Archives} Some bloggers (P1, P2, P4, P5, P10, P11) praised the Internet Archive for its efforts in archiving online content. Some participants (P2, P10, P11) also viewed the Internet Archive as a potential solution—or at least an additional safeguard—when used alongside other information infrastructure services. As one participant noted: \enquote{Yes, I think the Internet Archive for example is an admirable and incredibly important thing but now it is almost an emergency solution, since for a long time we did not have anything better} (P1). At the same time, this participant acknowledged limitations of the Internet Archive, pointing out that it is not easily searchable and that images are often missing. They added: \enquote{In the end it's a community or institutional solution that needs to be implemented} (P1). P13 added: \enquote{With something like the Internet Archive, I wouldn't feel safe. Especially not if I have anything to do with gender or postcolonialism. Since it’s foundation is based in the USA, the risk would be too great for me} (P13). Another participant raised concerns about the sustainability of online archives more broadly: \enquote{The problem with online archives is that they basically have the same disadvantage as the [blogging] platforms themselves, namely that they can disappear sometime, and then the contents are gone as well} (P5). P4 referred to the scholarly blog archive Rogue Scholar, praising the high quality of its support while noting that such intensive support is feasible only because of the archive’s current size. Participants also proposed conceptual ideas, such as a scholarly archival cloud (P1) or the use of Wikis or Wikidata due to their international reach (P3).

\paragraph{Platform Providers} Some participants, particularly those blogging on the scholarly platform de.hypotheses, reported relying on their platform to ensure the long-term preservation of their blogs. This also matches the answer of one participant that used the scholarly blogging platform SciLogs who also assumed that the publishing house already ensured the long-term preservation of blogs. The participant blogging on ScienceBlogs stated that they expect the publishing house behind the platform to make sure blog content is not lost but would also want an independent government funded institution to be involved.

\paragraph{Bloggers} Expectations toward the responsibility of bloggers were closely tied to their perceived abilities for long-term stability. Participants critiqued the expectations of individual responsibility for bloggers, due to time constraints, employment precarity, and a lack of technical expertise. P2 expressed a wish for person-independent infrastructures, so that the survival of a blog does not depend on bloggers. As they explained: 

\blockquote{Even with this simple WordPress system, the question was always: who will take care of continuously installing updates to the system, and who will ensure that the site remains accessible and reasonably stable after the next update, just as it is now online? And I don't think there's a patent solution for that. So, it can't be people. I myself am on a fixed-term contract. I simply don't know whether I will still be working in research in ten years' time. And if I'm no longer doing that, it's very unlikely that I'll still be the one providing updates. And that's why, if we want the content to be permanently accessible, we are dependent on infrastructures that are independent of individuals. (P2)} 

Other participants recognized that preservation cannot be delegated entirely to external institutions but remains partially dependent on the efforts of bloggers’. Some participants described themselves as responsible for preparing content for preservation by applying licenses and structuring metadata.

\section{Discussion}
\subsection{Dimensions of Infrastructure}
When asked about their requirements for blog preservation, most participants named or indirectly described existing infrastructure institutions and their services that already work towards preserving the scholarly record. Participants also named different actors to contribute to blog preservation and rejected the notion of a single responsible actor. This is reinforced by prior research, in which the proposed technical solutions \autocite{kalb2013, limani2017} have not been implemented, which likewise indicates that the preservation of scholarly blogs cannot be addressed through a single institutional or technical solution. On the basis of these accounts, we propose a decentralized approach that distributes copies and risk among diverse actors and ensures that no single actor holds exclusive responsibility for or control over scholarly blogs and the preservation of blogs is less vulnerable to cyberattacks, financial, and political instability. Because participants themselves pointed to existing institutions and services, we argue for extending their scope to include scholarly blogs, resulting in a decentralized information infrastructure for the long-term preservation of scholarly blogs that accommodates diverse positional, organizational, and political systems and activities. In this section, we ground this argument by aligning the implications of our findings with each dimension of infrastructure defined by Star and Ruhleder (\citeyear{star1996}). By doing that, we demonstrate that the circumstances for an infrastructure for the preservation of scholarly blogs already exist but are limited by social arrangements and require organizational efforts from infrastructure institutions, platform providers and bloggers.

\paragraph{Embeddedness} Star and Ruhleder (\citeyear{star1996}) describe infrastructure as being embedded within broader social, organizational, and technical arrangements rather than existing independently of them. Our findings demonstrate that the preservation of scholarly blogs is shaped by their surrounding structures and systems rather than by technical considerations alone. For example, some participants linked their lack of maintenance and preservation efforts to job insecurity due to fixed-term contracts and the insufficient institutional recognition of blogging. Additionally, while participants were asked about challenges concerning preservation, some of the challenges they named concerned blog maintenance. This notion suggests participants struggle with daily management of their blogs. Bloggers' missing efforts to preserve blogs are therefore a consequence of the way scholarly blogging is institutionally positioned outside established systems of scholarly communication. Preservation is delegated to bloggers instead of being distributed across dedicated infrastructure institutions. To address this disconnect, preservation activities should be embedded within existing institutional structure institutions, stronger formal recognition of blogging as scholarly output (e.g. through research funding organizations), support provided by libraries and Open Science offices, and the inclusion of blogs in institutional repositories. Rather than creating an entirely new infrastructure, our findings support embedding preservation responsibilities into existing socio-technical arrangements involving bloggers, infrastructure institutions, and platform providers. In this way, preservation becomes part of established scholarly communication practices instead of relying on individual effort. 

\paragraph{Transparency} Infrastructure should not need to be reinvented for each task but seemingly support them \autocite{star1996}. Our findings show that blog preservation currently lacks this characteristic. Participants repeatedly described uncertainty regarding preservation practices, metadata, licensing, and digital preservation more generally, demonstrating that preservation remains an explicit responsibility rather than an invisible infrastructural function. This suggests that the problem is not the absence of preservation technologies but the absence of preservation processes that operate transparently. Existing preservation infrastructures like repositories, libraries, Rogue Scholar, and the Internet Archive, already provide many required technical functions. Extending these services to scholarly blogs would therefore allow preservation to become an infrastructural property rather than an individual task for bloggers. For example, once a blog is registered with Rogue Scholar, DOI assignment, web archiving, and integration with the Internet Archive occur automatically. Such automation shifts preservation responsibilities away from individual bloggers and makes long-term preservation increasingly invisible in everyday blogging practice, reflecting Star and Ruhleder's (\citeyear{star1996}) notion of transparent infrastructure.

\paragraph{Reach or Scope} Infrastructure extends beyond single tasks or organizations and engages in different practices across communities, institutions, and time \autocite{star1996}. Our findings suggest that an infrastructure for blog preservation must likewise extend beyond individual bloggers and their disciplines, blogging platforms or preservation institutions. The requirements articulated by participants are diverse and include different stakeholders. Consequently, preservation cannot rely on institution-specific or platform-specific solutions. Instead, infrastructure must enable the integration of diverse blogs into a common, but decentralized preservation ecosystem. Achieving a decentralized approach, requires interoperable mechanisms for content exchange, such as standardized metadata, open licensing, and machine-readable feeds, which enable multiple infrastructure institutions to preserve and disseminate blogs across organizational and temporal boundaries. To aggregate diverse blogs we recommend that bloggers enable the reuse of their blogs available through the CC-BY license and provide blog posts via feeds \autocite{ochsner2026requirements} that enable the automated dissemination of blog content through standardized, machine-readable formats such as XML or JSON and are automatically generated by conventional blogging software \autocite{ochsner2026requirements, ochsner2025jdoc}.

\paragraph{Learned as Part of Membership} The use of an infrastructure requires acquired knowledge that is learned through the familiarization of the infrastructure through it's use by new members as they become members of a community of practice \autocite{star1996}. Communities of practice are informal, enduring groups of individuals who collectively learn and develop shared practices that reflect their mutual activities and social relationships over time \autocite{wenger1999}. Participants' limited knowledge about preservation, metadata, licensing, and rights management indicates that preservation has not yet become a normalized component of scholarly blogging. Rather than interpreting these knowledge gaps as individual deficiencies, our findings suggest that preservation practices have not yet been institutionalized within blogging communities. The infrastructure practices that need to be performed by bloggers consist of the communication with infrastructure institutions to opt into blog preservation, the provisions of open licenses on their blog, and the enrichment of blog feeds with basic metadata (e.g., blog name and short description, identifiers, editors and authors, license, and subject area \autocite{fenner2023_metadata}). Infrastructure institutions can support bloggers in acquiring the knowledge and skills required for infrastructure practices through targeted training programs. Training and resources could be embedded in existing information infrastructure services, such as academic libraries and repositories, archives, or offered directly through platform providers, such as hypotheses. At the same time, preservation infrastructures should minimize the amount of expertise required from bloggers through automation and user-friendly workflows. As preservation practices become integrated into routine blogging activities, they may gradually become taken for granted, allowing bloggers to engage with preservation without requiring an extensive technical knowledge.

\paragraph{Links with Conventions of Practice} Infrastructure influences and is influenced by communities of practice \autocite{star1996}. Communities of practice are informal, enduring groups of individuals who collectively learn and develop shared practices that reflect their mutual activities and social relationships over time \autocite{wenger1999}. By facilitating information sharing and interaction through blogrolls and comment sections, blogs foster dialogue, discussion, socialization and embed themselves within a community of practice \autocite{ochsner2025jdoc}. Participants did not only wish to preserve individual texts but often also comments, links, references, and, in some cases, the blogosphere. These findings reflect the social and interactive character of scholarly blogging that extends beyond individual posts to ongoing conversations and relationships between bloggers. Consequently, preservation should not treat blog posts as isolated documents but should aim to preserve their surrounding socio-technical context where feasible. Infrastructure institutions determine which elements of a blog remain accessible over time and decisions about what is preserved therefore shape the future scholarly record, illustrating the reciprocal relationship between infrastructure and scholarly communication practices. While bloggers are both contributors to preservation and infrastructure users, they also depend on technical and organizational support of infrastructure institutions which are embedded into existing conventions and regulations. Accordingly, preservation needs to be a distributed process in which bloggers, platform proverds, and infrastructure institutions contribute.

\paragraph{Embodiment of Standards} Infrastructure needs to incorporate standards that enable a transparent interoperability with other tools and systems, even amid conflicts \autocite{star1996}. Participants’ requests for metadata support shows that participants recognize interoperability as a prerequisite for the integration of blogs into infrastructures. Participants’ requests for stable identifiers, metadata support, and interoperability correspond to the embodiment of standards that already exist for more formal scholarly publications. To enable these services, an infrastructure for blog preservation needs to apply shared and interoperable technical and organizational standards, like metadata schemes, licensing frameworks, and interoperable data formats that collectively enable coordination and preservation across distributed stakeholders and systems. Efforts to make scholarly blogs citable have been made by the \href{https://datacite.org/community-groups/metadata-working-group/}{DataCite Metadata Working Group} that developed a proposal and a request for comments for the new resource type "BlogPost" \autocite{datacite} and \href{https://www.crossref.org/}{Crossref} that added the
subtype ‘blog’ to their Metadata deposit schema 5.5.0 \autocite{hendricks2026}. While standards for metadata, the interoperability of data and licensing already exist, our findings show that there is still a need for nuanced criteria for which content merits preservation, while considering that the preservation of content types such as links and comments is impacted by technical and legal constraints. While infrastructure institutions are generally well versed in these standards, bloggers also need to familiarize themselves with standards and practices. Lastly, considering, bloggers desire to retain control over preservation \autocite{ochsner2026iconf}, we suggests the adoption of an opt-in approach for the use of an infrastructure, to ensure that decisions about content preservation remain with bloggers \autocite{ochsner2026requirements}.

\paragraph{Built on an Installed Base} Infrastructure is not built but grown by layering new elements onto existing ones, rather than replacing them \autocite{star1996}. Our findings strongly support this principle. Rather than proposing entirely new preservation institutions, participants repeatedly referred to libraries, repositories, web archives, platform providers, Rogue Scholar, persistent identifiers, and existing metadata infrastructures as the basis for future preservation. These findings suggest that preserving scholarly blogs does not require building a separate infrastructure but extending the scope of existing preservation systems to include blogs as legitimate scholarly output. At the same time, the installed base also constrains innovation. Many repositories and preservation systems were originally designed for formal publications such as journal articles rather than dynamic web-based publications, meaning that organizational adjustments and new workflows will be necessary before blogs can be fully integrated. Accordingly, preservation should be understood as the gradual evolution of existing infrastructures rather than the creation of an entirely new one.

\paragraph{Becomes Visible Upon Breakdown} While infrastructure is often taken for granted, it becomes visible once it is not functional \autocite{star1996}. The general need for an infrastructure for the preservation of scholarly blogs becomes apparent, as blogs are disappearing due to platform instability, link decay, or risk disappearance due to political interference or loss of institutional support. Not having back-up mechanisms that preserve blog content reveal the necessity for an infrastructure for blog preservation. The data shows, how participants already notice the lack of infrastructure through the perceived threat of data loss due to recent political developments. A concern among many participants revolved around the perceived reliability of national government-funded institutions. The interviews for this study were conducted within the first 100 days of Donald Trump's second presidency as president of the United States (US). Following Trump's first presidency, the official website of the \href{https://www.whitehouse.gov/}{White House}, was altered and information on climate change made inaccessible \autocite{brugger2018}, followed by the loss of access of thousands of US-government web pages and datasets at the start of Trump's second presidency. The loss of access especially concerned topics on climate, diversity, and LGBTQ+ topics \autocite{carpenter2025}. Additionally, the far right political party Alternative für Deutschland (AFD, engl. alternative for Germany), received 20.8\% of votes in the German parliament elections of 2025 \autocite{bundeswahlleiterin2025}. As a result, many participants expressed that the US-American and German political situation at the time, negatively impacted their trust in national government structures and institutions. Therefore, participants expressed concerns about the vulnerability of government-led infrastructure institutions to data loss, interference, and censorship. Conversely, most participants still preferred a government-funded institution over commercial options. These findings support the need for a coalition of multiple stakeholders to safeguard scholarly output and freedom in light of ongoing political change. While scholarly blog preservation is not comprehensively ensured, responsibility and practices regarding scholarly blog preservation is already on a small scale distributed across multiple stakeholders \autocite{ochsner2025jdoc, ochsner2026iconf}, which is why we argue for decentralized preservation practices among national government funded information infrastructure institutions such as national and academic libraries and repositories, international digital archives and libraries such as the Internet Archive, CLOCKSS, and Rogue Scholar as well as platform providers. Especially, considering the principle of "Lots Of Copies Keep Stuff Safe" we argue for shared responsibility to mitigate the risk of information loss \autocite{clockss_home}.

The findings show that preservation is constrained by the coordination of distributed socio-technical relations. We therefore propose the establishment of a decentralized information infrastructure for the preservation of scholarly blogs by embedding preservation tasks into existing infrastructure institutions (e.g., libraries, repositories, and web archives) while considering existing social arrangements (e.g., between bloggers, their institutions and their platforms), organizational factors (e.g., services provides by infrastructure institutions) and technologies (e.g., existing software). We will make recommendations for the embedding of preservation efforts for blogs into these actors in the next section.

\subsection{Recommendations for Actors}
\paragraph{Libraries} Since preservation, curation, and the provision of stable access to knowledge have traditionally been core elements of libraries' institutional mission \autocite{borgman2007, nentwich2003}, libraries are well positioned to take responsibility for the long-term preservation of scholarly blogs. Most participants supported this view. Participants also identified discipline-specific information services and repositories as possible stakeholders in blog preservation. Because these services and repositories are often affiliated with academic libraries or research-performing institutions and already fulfill central functions in cooperative scientific information systems, they hold strong potential for integrating scholarly blogs. Repositories already provide persistent identifiers, quality-controlled metadata, and long-term citability for scientific content, and these functions could be extended to include blogs. Depositing scholarly blogs in several repositories (e.g., the institutional repository of the blogger's institution or academic library together with a subject-specific repository) would ensure that blog content is preserved by more than one actor, reducing the risk of information loss.

\paragraph{Web Archives} While some participants praised the web archiving efforts of the Internet Archive, others raised concerns about its safety and stability. In recent years, the Internet Archive has faced threats to its stability, including cyberattacks that temporarily restricted access to the Wayback Machine and legal disputes over copyright \autocite{brewster2024, freeland2024}. Some participants therefore agreed that the Internet Archive can not be solely responsible for safeguarding scholarly blogs. Preserving scholarly blogs through the Internet Archive alongside libraries and repositories would, however, ensure decentralization independently of government-funded institutions. Preserving individual blog posts in the Internet Archive also requires little technical knowledge and could be done by bloggers themselves. Rogue Scholar is a solution developed specifically for scholarly blogs and offers a model for how blog preservation could be organized. Unlike general web archives, it combines persistent identifiers, structured metadata, long-term archiving, and open licensing in a single workflow designed for scholarly blogging. Once bloggers register their blog, their posts receive DOIs and are archived automatically, which requires little technical effort. Because it is purpose-built for scholarly blogs, Rogue Scholar can serve as a central component of a decentralized preservation infrastructure. To support its' sustainability, Rogue Scholar is in the process of becoming a German non-profit organization to establish a more sustainable governance structure \autocite{fenner2025nonprofit}. One participant proposed the use of Wikidata, a free and collaborative repository that gathers structured data to support Wikipedia, Wikimedia Commons, and other wikis in the Wikimedia movement and their users \autocite{wikidata}; using Wikidata and wikis to preserve scholarly blogs comprehensively would, however, require a large community effort from bloggers themselves.

\paragraph{Platform Providers} Participants who blog on de.hypotheses see their platform as co-responsible for long-term preservation. Users of SciLogs and ScienceBlogs voiced a similar expectation but conceded that it is less realistic for commercial platforms. As a publicly funded project, de.hypotheses already assigns DOIs, requests ISSNs, and lists blogs in the OpenEdition catalog, acting as an intermediary between bloggers and infrastructure institutions. Hypotheses could build on this role by supplying blog content and metadata to libraries, repositories, or dedicated archives such as Rogue Scholar. For commercial platforms such as Medium or Substack, the role of profit-oriented companies in securing content, and their cooperation with public infrastructures, remains open. An important contribution they can make is to support export and interoperability through standardized metadata and machine-readable feeds, so that outside infrastructures can preserve the content even when the platform does not. Whether a platform is public or commercial, a transparent division of responsibility matters, since several participants relied on their platform for preservation.

\paragraph{Bloggers} Bloggers produce the content that is to be preserved and they are also the main users of any future infrastructure. For this reason they should be involved in building it. Several participants wanted to decide for themselves what is preserved and to keep control over their content, and an infrastructure designed without them would not meet this wish. If bloggers help design the preservation services, the services are more likely to fit how blogs are actually written and maintained. This can happen through co-design workshops, pilot tests with active bloggers, and consultation when metadata and licensing workflows are set up. Scholarly blogging already works as a community of practice, so existing networks and platforms such as de.hypotheses can be used to reach bloggers, gather their feedback, and share good practices. Working this way also builds preservation knowledge within the community instead of imposing it from outside, which speaks to the limited knowledge many participants reported. Handling blogs is part of a wider set of digital skills that should be developed further. Treating the preservation of scholarly blogs as part of good research practice would give bloggers a clearer reason to build these skills and would place blogs within the same framework of responsibility that already applies to research data and publications. Bloggers need to therefore do more than prepare their content: they need to help decide how the infrastructure operates.

\subsection{Limitations and Future Research}
This study has limitations. First, some bloggers that declined being part of this study expressed that they did not think they could contribute to long-term preservation, since they did not think they were educated enough on the topic. Therefore, some of the participants were bloggers who already had prior knowledge about infrastructures and preservation. However, the data revealed that even bloggers with prior experience on the topic still did not feel knowledgeable enough about possible preservation strategies. Second, both authors are involved in a project focused on the long-term preservation of scholarly blogs and therefore bring a pre-existing interest in ensuring their preservation. This positionality may have shaped the framing of the research questions and interpretation of the data. Accordingly, reflexive practices were employed throughout the analysis to mitigate potential bias. Third, as this study focused exclusively on the perspectives of a small sample of scholarly bloggers, the findings reflect only one stakeholder group within the broader preservation ecosystem. Information infrastructure institutions, platform providers, repositories, and web archives may hold different priorities, constraints, and understandings of responsibility. Fourth, the sample of participants consisted of many bloggers from the social sciences and humanities, even though we did not notice differences in the participants level of engagement or knowledge with preservation between disciplines. Lastly, this study was conducted with German participants and informed by the German federalized library system, which could potentially limit its generalizability to international applications. However, most of the actors that engage in preservation, do also exist in international contexts. 

While this study contributed to the existing research with novel findings, these findings could be expanded on by future research. Since this study was concerned with bloggers’ requirements for information infrastructures, future research could investigate good-practice solutions in information infrastructure institutions. Moreover, future research can build on this study by proposing a concrete practical implementation of a decentralized information infrastructure tasked with the preservation for scholarly blogs. Lastly, future research could investigate the long-term preservation of other types of scholarly outputs, such as microblogs, newsletters or podcasts.
\section{Conclusion}
The long-term preservation of scholarly blogs remains insufficiently ensured. To address this gap, we conducted 13 interviews with German scholarly bloggers to identify challenges bloggers encounter concerning preservation and the requirements they consider essential for an information infrastructure capable of supporting blog preservation. The results indicate that bloggers face challenges, consisting of limited technical expertise, a lack of institutional recognition and support, and ambiguity regarding responsibility for long-term preservation. Participants articulated clear expectations for an infrastructure that is trustworthy, transparent, and sustainable, while also respecting bloggers’ autonomy and intellectual property. Their limited technical knowledge extends to data handling, storage, rights management, and metadata, which reduces their willingness to engage in active long-term preservation. Fixed-term contracts, lack of institutional recognition, lack of support for blog production, and precarious hosting arrangements further impede sustainable preservation efforts. Participants requirements related to the preservation of relevant blog content were diverse. While some participants wished to save everything, conditional only on sufficient storage space, others argued for a more selective approach. Key technical and organizational needs included stable PIDs, metadata support, interoperability, and transparency regarding the conditions under which content is preserved. PIDs and metadata require coordination across bloggers, platform providers, and infrastructure institutions. Most participants preferred government-funded institutions over commercial options.

The requirements articulated by bloggers frame preservation not as a singular but as a distributed task. The scholarly blogosphere encompasses a heterogeneous set of stakeholders, with preservation efforts already being distributed among bloggers, platform providers, and information infrastructure institutions. It is unlikely that any single organization or infrastructure institution can assume sole responsibility for comprehensive preservation efforts. Findings show that no single stakeholder is perceived as capable of guaranteeing long-term preservation alone. Considering Star's and Ruhleder's (\citeyear{star1996}) dimensions of infrastructure we argue for extending the scope of existing decentralized infrastructure institutions to include preservation efforts for scholarly blogs. Such an infrastructure would distribute responsibility across multiple stakeholders, mitigate risks associated with political volatility, censorship, cyberattacks, and platform instability, and allow for different preservation routes. The findings also show that scholarly blogs are part of scholarly communication and cultural heritage, and their long-term preservation should not depend on individual labor alone. Building a distributed, interoperable, and community-supported infrastructure would not only secure long-term access to scholarly blogs but also strengthen their visibility, citability, and legitimacy within the broader scholarly communication ecosystem.

\section*{Funding}
The work of Catharina Ochsner is funded by the German Research Foundation (DFG) through the project Kooperative Informationsinfrastruktur f\"ur wissenschaftliche Blogs (Infra Wiss Blogs) (project number 528958385). Heinz Pampel was partly funded by the Einstein Center Digital Future (ECDF).

\printbibliography[heading=bibintoc]

\newpage
\renewcommand{\thepage}{\Alph{page}}
\setcounter{page}{1}

\end{document}